\documentstyle[epsf,rotate]{ichep}
\newcommand{\gev}{{\rm GeV}}
\newcommand{\mev}{{\rm MeV}}

\newcommand{\eqn}[1]{(\ref{#1})}

\begin{document}

\title{
{\normalsize\rm
\rightline{TUM--T31--77/94}
\rightline{September 1994}
\ \\
\ \\
\ \\}
Some Prospects for the Determination of the Unitarity Triangle
before the LHC Era*}
\collab{\normalsize\bf
Note to the preprint reader: Due to limitations of the plotting program \\
the axis labels in figs.\ \ref{fig:utriag} and \ref{fig:sinesab} are
missing the bars on $\varrho$ and $\eta$.}

\author{Markus E.\ Lautenbacher$^{\dag}$}

\affil{
Physik Department, Technische Universit\"at M\"unchen, \\
D--85748 Garching, Germany.
}

\abstract{
Anticipating improved determinations of $m_t$, $|V_{cb}|$,
$|V_{ub}/V_{cb}|$, $B_K$ and $F_B \sqrt{B_B}$ in the next five years we
make an excursion into the future in order to find a possible picture of
the unitarity triangle, of quark mixing and of CP violation around the
year~2000. We then analyse what impact on this picture will result from
the measurements of the four possibly cleanest quantities: $BR(K^+ \to
\pi^+ \nu \bar\nu)$, $x_d/x_s$, $\sin(2\alpha)$ and $\sin(2\beta)$.  In
the course of our investigations we extend the analysis of the
unitarity triangle beyond the leading order in the Wolfenstein
parameter $\lambda$. \\
We will also shortly present the status of direct CP violation in $K
\rightarrow \pi\pi$ and $K_{\rm L} \rightarrow \pi^0 e^+ e^-$.
}

\twocolumn[\maketitle]

\fnm{7}{Invited talk given at ICHEP '94, Glasgow, July 1994.}

\fnm{1}{E-mail: {\tt lauten@feynman.t30.physik.tu-muenchen.de}}

\section{CKM Matrix and Unitarity Triangle}
   \label{sec:ckmmatrix}
\subsection{Wolfenstein Parametrization Beyond Leading Order}
In the Standard Model (SM) with three fermion generations, CP violation
arises from a single phase in the unitary $3 \times 3$
Cabibbo-Kobayashi-Maskawa (CKM) matrix.  For phenomenological
applications it is useful to expand each element of the CKM matrix as a
power series in the small parameter $\lambda = |V_{us}| = 0.22$. For
the leading order in $\lambda$ the result is \cite{wolfenstein:83}
\begin{equation}
V_{\rm CKM} = \left(
\begin{array}{ccc}
1 - \frac{\lambda^2}{2} & \lambda & A \lambda^3 (\varrho - i \eta) \\
-\lambda & 1 - \frac{\lambda^2}{2} & A \lambda^2  \\
A \lambda^3 (1 - \varrho - i \eta) & -A \lambda^2 & 1
\end{array}
\right) + {\cal O}(\lambda^4)
\label{eq:wolf}
\end{equation}
This parametrization being an expansion in $\lambda$ respects unitarity
of the CKM matrix only approximately up to terms of order ${\cal
O}(\lambda^5)$. With e.g.\ LHC expected to test unitarity to a very high
precision one has to extend the expansion \eqn{eq:wolf} to higher order terms
in $\lambda$. As always with next-to-leading order expansions the
definition of higher terms is not unique. A particularly nice form is to
relate the parameters $(\lambda, A, \varrho, \eta)$ of the approximate
Wolfenstein parametrization to the parameters $s_{ij}$ and $\delta$ of
the fully unitary standard parametrization \cite{particledata:92} of
the CKM matrix through \cite{burasetal:94b}
\begin{equation}
s_{12} \equiv \lambda      \quad
s_{23} \equiv A \lambda^2  \quad
s_{13} \, {\rm e}^{-i \delta} \equiv A \lambda^3 (\varrho - i \eta) \; .
\label{eq:nlodef}
\end{equation}

\subsection{Unitarity Triangle Beyond Leading Order}
The unitarity of the CKM-matrix provides us with several relations
of which
\begin{equation}
V_{ud}^{}V_{ub}^* + V_{cd}^{}V_{cb}^* + V_{td}^{}V_{tb}^* =0
\label{eq:unitarity}
\end{equation}
is the most useful one.  In the complex plane the relation
\eqn{eq:unitarity} can be represented as a triangle, the so-called
``unitarity--triangle'' (UT).  Phenomenologically this triangle is very
interesting as it involves simultaneously the CKM elements $V_{ub}$,
$V_{cb}$ and $V_{td}$ which are under extensive discussion at present.

In the usual analyses of the unitarity triangle only terms ${\cal
O}(\lambda^3)$ are kept in \eqn{eq:unitarity}.  Including the
next-to-leading terms by keeping ${\cal O}(\lambda^5)$ corrections and
rescaling all terms in \eqn{eq:unitarity} by $|V_{cd}^{} V_{cb}^*| = A
\lambda^3 + {\cal O}(\lambda^7)$ we find

\begin{equation}
 \frac{1}{A\lambda^3}V_{ud}^{}V_{ub}^* = \bar\varrho + i \, \bar\eta
\quad
 \frac{1}{A\lambda^3}V_{td}^{}V_{tb}^* = 1-(\bar\varrho + i \, \bar\eta)
\label{eq:rescaled}
\end{equation}
with $\bar\varrho$ and $\bar\eta$ defined by \cite{burasetal:94b}
\begin{equation}
\bar\varrho = \varrho \, (1-\frac{\lambda^2}{2})
\qquad
\bar\eta = \eta \, (1-\frac{\lambda^2}{2}).
\label{eq:bardef}
\end{equation}

\begin{figure}[htb]
\centerline{
\epsfysize=1in
\epsffile{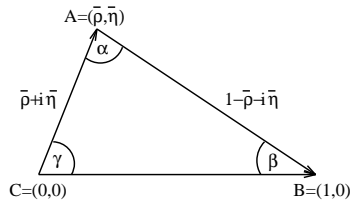}
}
\caption[]{
Unitarity triangle in the complex $(\bar\varrho,\bar\eta)$ plane.
\label{fig:triangle}}
\end{figure}

Thus we can represent \eqn{eq:unitarity} as a triangle, the UT, in the
complex $(\bar\varrho,\bar\eta)$ plane. This is  shown in
fig.~\ref{fig:triangle}.

We observe that beyond the leading order in $\lambda$ the point `A' {\it
does not} correspond to  $(\varrho,\eta)$ but to
$(\bar\varrho,\bar\eta)$. Clearly within 3\% accuracy
$\bar\varrho=\varrho$ and $\bar\eta=\eta$.  Yet in the distant future
the accuracy of experimental results and theoretical calculations may
improve considerably so that the more accurate formulation given here
will be appropriate. For instance the experiments at LHC should measure
$ \sin(2\beta) $ to an accuracy of (2--3)\% \cite{camilleri:93}.

With fig.~\ref{fig:triangle} it is then a matter of simple trigonometry
to calculate $\sin(2\phi_i$) in terms of $(\bar\varrho,\bar\eta)$ and
vice versa.

In sects.\ \ref{sec:presentday}--\ref{sec:sin2ab} we will now summarize
the phenomenological analysis of the UT presented in
ref.\ \cite{burasetal:94b}.

\section{The UT from Present Day Experiments}
   \label{sec:presentday}
\subsection{Tree Level B-Decays}
Measurements of tree level B-decays can be used to derive the CKM
elements $|V_{cb}|$, $|V_{ub}/V_{cb}|$. This then allows to determine
\begin{equation}
R_b \equiv \frac{|V_{ud}^{}V^*_{ub}|}{|V_{cd}^{}V^*_{cb}|}
= \sqrt{\bar\varrho^2 +\bar\eta^2}
= (1-\frac{\lambda^2}{2})\frac{1}{\lambda}
\left| \frac{V_{ub}}{V_{cb}} \right|
\label{eq:Rb}
\end{equation}
which represents a circle centered around $(0,0)$ in the complex
$(\bar\varrho,\bar\eta)$ plane.  Thus $R_b$ is simply the length
$\overline{AC}$ in the rescaled UT of fig.~\ref{fig:triangle}.

\subsection{Indirect CP Violation}
The usual box diagram calculation together with the experimental value
for $\varepsilon_K$ from $K^0$-$\bar K^0$ mixing specifies a hyperbola
in the $(\bar\varrho, \bar\eta)$ plane with $\bar\eta > 0$
\cite{harrisrosner:92,dibdunietzgilman:90}:
\begin{equation}
\bar\eta \left[(1-\bar\varrho) A^2 \eta_2 S(x_t)
+ \left[ \eta_3 S(x_c,x_t) - \eta_1 x_c \right] \frac{1}{\lambda^4} \right]
= \frac{0.223}{A^2 B_K}
\label{eq:epsK}
\end{equation}
Here $S(x_i)$, $S(x_i,x_j)$, $x_i = m^2_i/M^2_W$ are the Inami-Lim
functions, $B_K$ is the renormalization group invariant
non-perturbative parameter describing the size of $<\bar{K}^0|
(\bar s d)_{V-A}(\bar s d)_{V-A}|K^0>$ and $\eta_1 = 1.1$
\cite{herrlichnierste:93}, $\eta_2 = 0.57$ \cite{burasjaminweisz:90},
$\eta_3=0.36$
\cite{kaufmanetal:88}--\nocite{buchallaetal:90,dattaetal:90}\cite{flynn:90}
represent QCD corrections to the box diagrams.

\subsection{$B^0$-$\bar B^0$ Mixing}
The experimental knowledge of the $B^0_d-\bar B^0_d$ mixing described by
the parameter $x_d = \Delta M/\Gamma_B$ determines $|V_{td}|$.
This then specifies via
\begin{equation}
R_t \equiv \frac{|V_{td}^{}V^*_{tb}|}{|V_{cd}^{}V^*_{cb}|} =
 \sqrt{(1-\bar\varrho)^2 +\bar\eta^2}
=\frac{1}{\lambda} \left| \frac{V_{td}}{V_{cb}} \right|
\label{eq:Rt}
\end{equation}
a circle centered around $(1,0)$ in the complex $(\bar\varrho,\bar\eta)$ plane.
Here $R_t$ is simply the length $\overline{AB}$ in the rescaled UT of
fig.~\ref{fig:triangle}.

All the QCD corrections to $\varepsilon_K$, $B^0-\bar B^0$ mixing and
$BR(K^+ \rightarrow \pi^+ \nu\bar\nu)$ used here include except for
$\eta_3$ the next-to-leading order. Hence, in all formulae of this
paper $m_t$ corresponds to the running top quark mass in the
$\overline{MS}$ scheme evaluated at $m_t$ i.e. $m_t = \overline{m}_t(m_t)$.
The physical top quark mass as the pole of the renormalized propagator
is for the range of $m_t$ considered here by $(7 \pm 1)\,\gev$ higher
than $m_t$.

\begin{figure}[htb]
\centerline{
\epsfysize=2in
\rotate[r]{
\epsffile{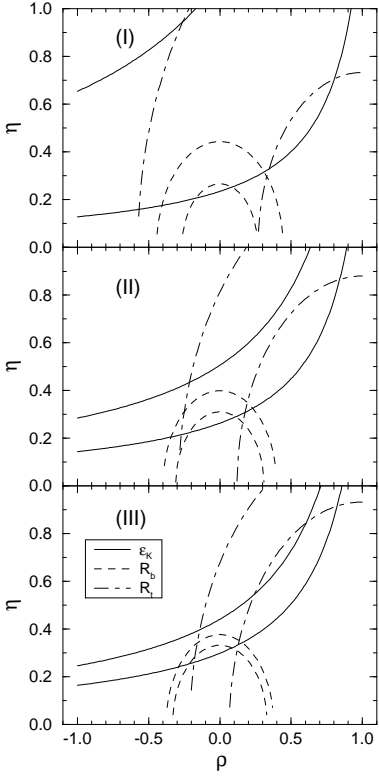}
}
}
\caption[]{
Unitarity triangle in the $(\bar\varrho,\bar\eta)$ plane determined by
$\varepsilon_{K}$, $|V_{cb}|$, $|V_{ub}/V_{cb}|$ and $x_d$ using ranges
(I)--(III) as of tab.\ \ref{tab:inputranges1}.
\label{fig:utriag}}
\end{figure}

Using eqs.\ \eqn{eq:Rb}--\eqn{eq:Rt} together with present day and
envisioned future ranges of input parameters as of
tab.\ \ref{tab:inputranges1} and $\Lambda_{\overline{MS}} = 300\,\mev$,
$m_c = 1.3\,\gev$, one can determine the allowed ranges for the upper
corner `A' of the UT and make predictions for various quantities. The
result is shown in fig.\ \ref{fig:utriag} and
tab.\ \ref{tab:outputranges1}, respectively.

\begin{table}[b]
\caption[]{Present day and envisioned ranges of input parameters for
the determination of the UT from tree level B-decays, indirect CP
violation and $B^0$-$\bar B^0$ mixing.}
\begin{center}
\begin{tabular}{c||c|c|c}
Parameter Range & (I) & (II) & (III) \\
      & (1994) & ($\sim$ 1997) & ($\sim$ 2000) \\
\hline
\hline
$|V_{cb}|$        & $0.038 \pm 0.004$ & $0.040 \pm 0.002$ & $0.040 \pm 0.001$
\\
$|V_{ub}/V_{cb}|$ & $0.08 \pm 0.02$ & $0.08 \pm 0.01$ & $0.08 \pm 0.005$ \\
$B_K$ & $0.7 \pm 0.2$ & $0.75 \pm 0.07$ & $0.75 \pm 0.05$ \\
$F_{B_d} \sqrt{B_{B_d}}$\,[\mev] & $200 \pm 30$ & $185 \pm 15$ & $185 \pm 10$
\\
$x_d$ & $0.72 \pm 0.08$ & $0.72 \pm 0.04$ & $0.72 \pm 0.04$ \\
$m_t$\,[\gev] & $165 \pm 15$ & $170 \pm 7$ & $170 \pm 5$ \\
\end{tabular}
\end{center}
\label{tab:inputranges1}
\end{table}

\begin{table}[htb]
\caption[]{Predictions for various quantities using input parameters as of
tab.\ \ref{tab:inputranges1}}
\begin{center}
\begin{tabular}{c||c|c|c}
Parameter Range & (I) & (II) & (III) \\
\hline
\hline
$\sin(2 \alpha)$ & $0.17 \pm 0.84$ & $0.35 \pm 0.65$ & $0.50 \pm 0.49$ \\
$\sin(2 \beta)$  & $0.59 \pm 0.21$ & $0.60 \pm 0.14$ & $0.61 \pm 0.09$ \\
$\sin(2 \gamma)$ & $0 \pm 1$ & $0 \pm 0.88$ & $0 \pm 0.68$ \\
$|V_{td}| \times 10^3$ & $9.4 \pm 2.5$ & $9.5 \pm 1.4$ & $9.4 \pm 1.0$ \\
$x_s$ & $16.0 \pm 8.3$ & $13.4 \pm 4.3$ & $12.9 \pm 2.8$ \\
$BR(K^+ \rightarrow \pi^+ \nu\bar\nu) \times 10^3$ & $1.04 \pm 0.42$ &
$1.07 \pm 0.24$ & $1.03 \pm 0.15$
\end{tabular}
\end{center}
\label{tab:outputranges1}
\end{table}

Looking at tab. \ref{tab:outputranges1} one sees that by the year 2000
one can expect predictions for $\sin(2 \beta)$, $|V_{td}|$, $BR(K^+
\rightarrow \pi^+ \nu\bar\nu)$ good to $\pm$(10--15)\% and for $x_s$ up to
$\pm$20\%. Thus a measurement of $BR(K^+ \rightarrow \pi^+ \nu\bar\nu)$ or
$x_s$ at the level of $\pm$10\% could serve as a possible test of the
corresponding SM predictions. Huge uncertainties for predicting $\sin(2
\alpha)$ and $\sin(2 \gamma)$ remain however, even with improved input
in the future. Turning the argument around, this signals that a
measurement of one of these angles would allow to put stringent
constraints on some of the input parameters of tab.\ \ref{tab:inputranges1},
e.g.\ the non-perturbative ones $B_K$ and $F_{B_d} \sqrt{B_{B_d}}$.

\section{The UT from CP Violating B-Asymmetries}
   \label{sec:Basym}
Measuring the CP-asymmetries in neutral B-decays will give the
definitive answer whether the CKM description of CP violation is
correct. Assuming that this is in fact the case, we want to investigate
the impact of the measurements of $\sin(2\phi_i)$ on the determination
of the unitarity triangle.  Since in the rescaled triangle of
fig.~\ref{fig:triangle} one side is known, it suffices to measure two
angles to determine the triangle completely.

With the CP-asymmetries simply given by
\begin{eqnarray}
A_{CP}(B^0 \rightarrow \psi K_{\rm S}) &=& -\sin(2 \beta) \,
\frac{x_d}{1+x_d^2}
\label{eq:Asin2b}
\\
A_{CP}(B^0 \rightarrow \pi^+ \pi^-) &=& -\sin(2 \alpha) \, \frac{x_d}{1+x_d^2}
\label{eq:Asin2a}
\end{eqnarray}
one can determine $\sin(2 \beta)$ without any theoretical uncertainties
from measuring the CP-asymmetry in $B^0 \rightarrow \psi K_{\rm S}$,
while for $\sin(2 \alpha)$ the measurement of several other channels is
required in order to remove the penguin contributions.

\begin{figure}[htb]
\centerline{
\epsfysize=2in
\rotate[r]{
\epsffile{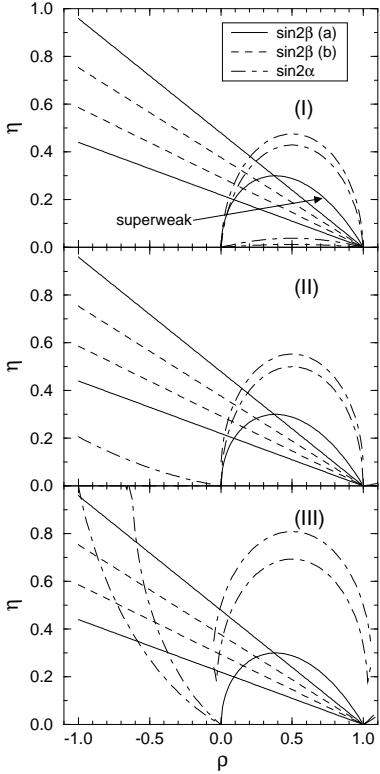}
}
}
\caption[]{
Determination of the unitarity triangle in the $(\bar\varrho,\bar\eta)$
plane by measuring $\sin(2\beta)$ and $\sin(2\alpha)$ as of
eqs.~\eqn{eq:sin2branges} and \eqn{eq:sin2aranges}, respectively.  For
$\sin(2\alpha)$ we always find two solutions in $(\bar\varrho,
\bar\eta)$ and for $\sin(2\beta)$ we only use the solution consistent
with $|V_{ub}/V_{cb} |\leq 0.1$.
\label{fig:sinesab}}
\end{figure}

Assuming a measurement of $\sin(2 \beta)$ and $\sin(2 \alpha)$ to give
\begin{eqnarray}
\sin(2 \beta) &=& \left\{
\begin{array}{ll}
0.60 \pm 0.18 & \mbox{(a) HERA-B \cite{albrechtetal:92}} \\
0.60 \pm 0.06 & \mbox{(b) SLAC \cite{babar:93}}
\end{array}
\right.
\label{eq:sin2branges}
\\
\sin(2 \alpha) &=& \left\{
\begin{array}{rl}
-0.20 \pm 0.10 & \mbox{(I)}   \\
 0.10 \pm 0.10 & \mbox{(II) SLAC \cite{babar:93}} \\
 0.70 \pm 0.10 & \mbox{(III)}
\end{array}
\right.
\label{eq:sin2aranges}
\end{eqnarray}
with the errors expected from different experiments indicated, one can
again determine the UT in $(\bar\varrho, \bar\eta)$ space. The result is
shown in fig.\ \ref{fig:sinesab}. Here the solid line labeled `superweak'
reflects the implicit relation holding between $\bar\varrho$ and $\bar\eta$
in the superweak scenario where $\sin(2 \beta) = -\sin(2 \alpha)$.

Comparing figs.\ \ref{fig:utriag} and \ref{fig:sinesab} it is obvious
that a combined measurement of $\sin(2 \beta)$ and $\sin(2 \alpha)$ at
the expected precision will have a large impact on the determination of
the UT and CKM parameters (For a discussion of $\sin(2 \beta)$ and
$\sin(2 \gamma)$ see ref.\ \cite{burasetal:94b}.). E.g.\ using $\sin(2
\beta) = 0.6 \pm 0.06$, $\sin(2 \alpha) = 0.1 \pm 0.1$ and range~(II) of
tab.\ \ref{tab:inputranges1} for $|V_{cb}|$, $x_d$ and $m_t$ one
obtains $\sin(2 \gamma) = 0.54 \pm 0.12$, $|V_{td}| = (8.8 \pm 0.4)
\times 10^{-3}$, $x_s = 16.3 \pm 1.3$ and $BR(K^+ \rightarrow \pi^+
\nu\bar\nu) = (1.01 \pm 0.11) \times 10^{-10}$.

The ability to make predictions for $|V_{td}|$, $x_s$ and $BR(K^+
\rightarrow \pi^+ \nu\bar\nu)$ to an accuracy of $\pm$(5--10)\% stems
from the absent or small theoretical uncertainties in
eqs.\ \eqn{eq:Asin2b} and \eqn{eq:Asin2a}, as well as from the expected
high precision for the measurement of CP violating B-asymmetries.
However, this predictive power can only be achieved through a
measurement of both $\sin(2 \beta)$ and $\sin(2 \alpha)$. Finally, we
note that the predictions resulting from a measurement of CP
violating B-asymmetries are generally more precise than those using
$\varepsilon_K$, $x_d$, $|V_{cb}|$ and $|V_{ub}/V_{cb}|$ as input
data.

\section{$\sin(2 \beta)$, $\sin(2 \alpha)$ from Indirect CP Violation
and $B^0$-$\bar B^0$ Mixing versus a Direct Measurement}
   \label{sec:sin2ab}
\begin{figure}[htb]
\centerline{
\epsfysize=4in
\epsffile{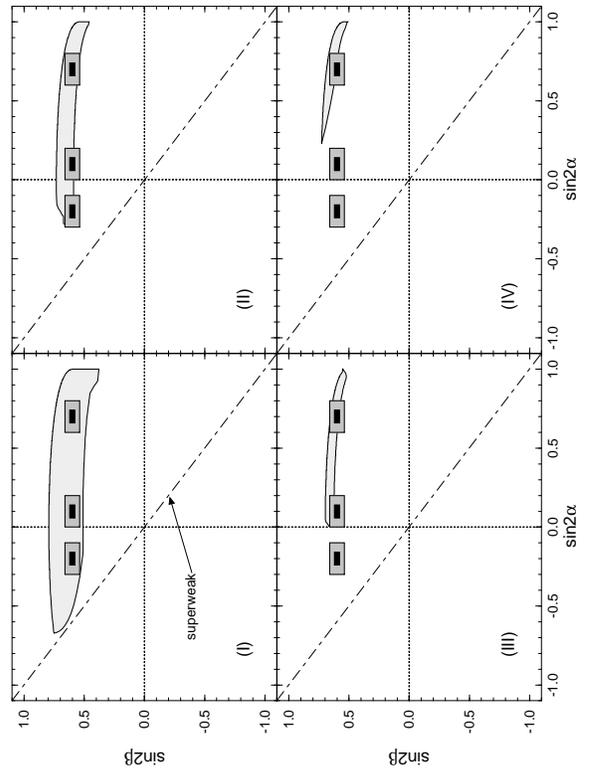}
}
\caption[]{
$\sin(2\alpha)$ versus $\sin(2\beta)$ plot corresponding to the
parameter ranges (I)--(IV) as of tab.\ \ref{tab:inputranges1} and
eq.\ \eqn{eq:rangeIV}. The dark shaded rectangles are given by
eqs.\ \eqn{eq:sin2aranges} and \eqn{eq:sin2branges}\,(b). The black
rectangles illustrate the accuracy of future LHC measurements.
\label{fig:sin2bvs2a}}
\end{figure}

It is useful to combine the results of sects.\ \ref{sec:presentday} and
\ref{sec:Basym} by making the customary $\sin(2\beta)$ versus
$\sin(2\alpha)$ plot \cite{nir:74}. This plot demonstrates very clearly
the correlation between $\sin(2\alpha)$ and $\sin(2\beta)$. The allowed
ranges for $\sin(2\alpha)$ and $\sin(2\beta)$ corresponding to the
choices of parameters in tab.\ \ref{tab:inputranges1} are shown in
fig.~\ref{fig:sin2bvs2a} together with the results of the independent
measurements of $\sin(2\beta) = 0.60 \pm 0.06$ and $\sin(2\alpha)$
given by \eqn{eq:sin2aranges}. The latter are represented by dark
shaded rectangles. The black rectangles illustrate the accuracy of
future LHC measurements ($\Delta\sin(2\alpha) = \pm 0.04$,
$\Delta\sin(2\beta) = \pm 0.02$) \cite{camilleri:93}. We also show the
results of an analysis in which the accuracy of various parameters is
as in range (II) of tab.\ \ref{tab:inputranges1} but with the central
values modified. Parameter range~(IV) is given by
\begin{equation}
\begin{array}{ll}
\left| V_{cb} \right| = 0.038 \pm 0.002 & |V_{ub}/V_{cb}| = 0.08 \pm 0.01 \\
B_K = 0.70 \pm 0.07 & \sqrt{B_{B_d}} F_{B_d} = (185 \pm 15)\,\mev \\
x_d = 0.72 \pm 0.04 & m_t = (165 \pm 7)\,\gev \\
\end{array}
\label{eq:rangeIV}
\end{equation}
In addition we show the prediction of superweak theories which in this plot
is represented by a straight line.

There are several interesting features visible on this plot:
First, the impact of the direct measurements of $\sin(2\beta)$ and
$\sin(2\alpha)$ is clearly visible in fig.\ \ref{fig:sin2bvs2a}. \\
Next, in cases (III) and (IV) we have examples where the measurements of
$\sin(2\alpha)$ are incompatible with the predictions coming from
$\varepsilon_{K}$ and $B^0 - \bar{B^0}$ mixing. This would be a signal
for physics beyond the standard model. The measurement of
$\sin(2\alpha)$ is essential for this. \\
Furthermore, case (IV) shows that for a special choice of parameters
the predictions for the asymmetries coming from $\varepsilon_{K}$, $B^0
- \bar{B^0}$ mixing, $|V_{cb}|$ and $|V_{ub}/V_{cb}|$ can be quite
accurate when these four constraints can only be satisfied
simultaneously in a small area of the $(\bar\varrho, \bar\eta)$~space.
Decreasing $|V_{cb}|$, $|V_{ub}/V_{cb}|$ and $m_t$ and increasing $F_B$
would make the allowed region in the case (IV) even smaller. \\
Finally, we also observe that the future measurements of B-asymmetries
and the improved ranges for the parameters relevant for
$\varepsilon_{K}$ and $B^0 - \bar{B^0}$ mixing will probably allow to
rule out the superweak models. This was also already indicated by
fig.\ \ref{fig:sinesab}\,(III).

\section{Direct CP Violation in $K \rightarrow \pi\pi$ and $K_{\rm L}
\rightarrow \pi^0 e^+ e^-$}
   \label{sec:CPviolK}
$Re(\varepsilon'/\varepsilon)$ measures the ratio of direct to indirect
CP violation in $K \rightarrow \pi\pi$ decays. The short distance QCD
corrections to $\varepsilon'/\varepsilon$ have been calculated at the
next-to-leading order level \cite{burasetal:92d,ciuchini:92}. The
result of these analyses can be summarized in an analytic formula for
$Re(\varepsilon'/\varepsilon)$ as a function of $m_t$,
$\Lambda_{\overline{MS}}$, $m_s$, hadronic matrix element parameters
$B_6$, $B_8$ and CKM elements \cite{buraslauten:93}. A simplified
version of this formula is given by
\begin{equation}
Re(\varepsilon'/\varepsilon) = 12 \cdot 10^{-4}
\left[ \frac{\eta \lambda^5 A^2}{1.7 \cdot 10^{-4}} \right]
\left[ \frac{150\,\mev}{m_s(m_c)} \right]^2
\left[ B_6 - Z(x_t) B_8 \right],
\label{eq:epe}
\end{equation}
where $Z(x_t) = 0.175 \cdot x_t^{0.93}$. Eq.\ \eqn{eq:epe} clearly
shows that in the SM $\varepsilon'/\varepsilon$ is governed by QCD
($B_6$) and electroweak ($B_8$) penguin contributions. For $m_t = (170
\pm 10)\,\gev$ and using $\varepsilon_K$-analysis to determine $\eta$
one finds \cite{burasetal:92d} $Re(\varepsilon'/\varepsilon) = (6 \pm
4) \times 10^{-4}$ for $B_6 = B_8 \approx 1$ (lattice, $1/N$
expansion).  For $B_6 \approx 2$, $B_8 \approx 1$ (QCD penguin
domination) values as high as $Re(\varepsilon'/\varepsilon) = (15 \pm
5) \times 10^{-4}$ are possible. Thus the remaining theoretical
uncertainty stemming from hadronic parameters somehow resembles the
still existing experimental discrepancy between E731
$Re(\varepsilon'/\varepsilon) = (7.4 \pm 5.9) \times 10^{-4}$
\cite{gibbons:93} and NA31 $Re(\varepsilon'/\varepsilon) = (23 \pm 7)
\times 10^{-4}$ \cite{barr:93}.

For the decay $K_{\rm L} \rightarrow \pi^0 e^+ e^-$ a recent
next-to-leading order analysis \cite{burasetal:94a} of the directly CP
violating contribution indicates this part of the amplitude to be the
dominant one. One obtains $BR(K_{\rm L} \rightarrow \pi^0 e^+ e^-)_{\rm
dir} = (6 \pm 3)\times 10^{-12}$ \cite{burasetal:94a} and
$BR(K_{\rm L} \rightarrow \pi^0 e^+ e^-)_{\rm indir} \le 1.6 \times
10^{-12}$, $BR(K_{\rm L} \rightarrow \pi^0 e^+ e^-)_{\rm cons} = (1.0
\pm 0.8)\times 10^{-12}$ for the indirectly CP violating and CP
conserving contributions, respectively \cite{cohenetal:93,pich:93}.
The present experimental bound is $BR(K_{\rm L} \rightarrow \pi^0 e^+
e^-) \le 4.3 \times 10^{-9}$ \cite{harrisetal:93}.

\section{Summary and Conclusions}
   \label{sec:summary}
We have shown that in order to compete with the accuracy expected from
LHC for the determination of the UT one needs to extend the usual
Wolfenstein parametrization of the CKM matrix to the next-to-leading
order in the expansion in terms of $\lambda$. To this end we have
proposed a form of the next-to-leading order expansion for which the UT
at next-to-leading order in $\lambda$ nicely resembles the UT in
leading order when coordinates are expressed in
$(\bar\varrho,\bar\eta)$ instead of the usual ones $(\varrho,\eta)$.

Our analysis investigated how well the UT can possibly be determined
around the year 2000 from data on $\varepsilon_K$, $B^0$--$\bar B^0$
mixing, $|V_{cb}|$ and $|V_{ub}/V_{cb}|$. We have found that along this line
it will be possible to make predictions for $|V_{td}|$, $\sin(2 \beta)$
and $BR(K^+ \rightarrow \pi^+ \nu\bar\nu)$ up to an error of
$\pm$(10--15)\%.  However, for $x_s$ and $\sin(2 \alpha)$, $\sin(2
\gamma)$ there will remain sizeable/huge uncertainties, respectively.
This results from theoretical uncertainties being present already in the
determination of some of the input parameters of this approach.

On the other hand, the future determination of $\sin(2 \alpha)$ and
$\sin(2 \beta)$ from CP violating B-asymmetries at HERA-B, SLAC, KEK
being (almost) free of theoretical uncertainties turns out to have an
impressive impact on our knowledge of the UT. Along this line it will
e.g.\ be possible to predict $|V_{td}|$, $x_s$ and $BR(K^+ \rightarrow
\pi^+ \nu\bar\nu)$ up to an error of $\pm$(5--10)\%. Future LHC
B-physics experiments around the year 2005 will refine these studies as
evident from fig.~\ref{fig:sin2bvs2a} and ref.~\cite{camilleri:93}

Any discrepancy found between the indirect determination of $\sin(2
\alpha)$, $\sin(2 \beta)$ from \{$\varepsilon_K$, $x_d$, $|V_{cb}|$,
$|V_{ub}/V_{cb}|$\} and a direct measurement in CP violating
B-asymmetries would signal new physics beyond the SM.

Finally, we shortly summarized the status of direct CP violation in
$K \rightarrow \pi\pi$ where for $\varepsilon'/\varepsilon$ both
experiment and the non-perturbative part of theory need some
improvements. While direct CP violation is known to give only a small
contribution to the whole amplitude in $K \rightarrow \pi\pi$, our
recent analysis of the direct CP violating part in the decay $K_{\rm L}
\rightarrow \pi^0 e^+ e^-$ indicates that there this contribution seems
to be the dominant one.

\medskip
\leftline{\bf Acknowledgments}
The author would like to thank A.\,J.\ Buras and G.\ Ostermaier for
the most pleasant collaboration on the work presented here.


\Bibliography{10}
\bibitem{wolfenstein:83}
{\sc L.~Wolfenstein},
\newblock {\em Phys.~Rev.~Lett.} {\bf 51} (1983) 1945.

\bibitem{particledata:92}
{\sc {Particle~Data~Group}},
\newblock {\em Phys.~Rev.} {\bf D~45} (1992) No.11~part~II.

\bibitem{burasetal:94b}
{\sc A.~J. Buras}, {\sc M.~E. Lautenbacher}, and {\sc G.~Ostermaier},
\newblock {\em Phys. Rev.} {\bf D} (1994) {\rm (in print), hep-ph 9403384}.

\bibitem{camilleri:93}
{\sc L.~Camilleri},
\newblock {\em CERN preprint} {\bf CERN-PPE/93-159}.

\bibitem{harrisrosner:92}
{\sc G.~R. Harris} and {\sc J.~L. Rosner},
\newblock {\em Phys.~Rev.} {\bf D~45} (1992) 946.

\bibitem{dibdunietzgilman:90}
{\sc C.~O. Dib}, {\sc I.~Dunietz}, {\sc F.~J. Gilman}, and {\sc Y.~Nir},
\newblock {\em Phys. Rev.} {\bf D~41} (1990) 1522.

\bibitem{herrlichnierste:93}
{\sc S.~Herrlich} and {\sc U.~Nierste},
\newblock {\em Nucl.~Phys.} {\bf B419} (1994) 292.

\bibitem{burasjaminweisz:90}
{\sc A.~J. Buras}, {\sc M.~Jamin}, and {\sc P.~H. Weisz},
\newblock {\em Nucl.~Phys.} {\bf B347} (1990) 491.

\bibitem{kaufmanetal:88}
{\sc W.~A. Kaufman}, {\sc H.~Steger}, and {\sc Y.~P. Yao},
\newblock {\em Mod.~Phys.~Lett.} {\bf A3} (1988) 1479.

\bibitem{buchallaetal:90}
{\sc G.~Buchalla}, {\sc A.~J. Buras}, and {\sc M.~K. Harlander},
\newblock {\em Nucl.~Phys.} {\bf B337} (1990) 313.

\bibitem{dattaetal:90}
{\sc A.~Datta}, {\sc J.~Fr{\"o}hlich}, and {\sc E.~A. Paschos},
\newblock {\em Zeitschr.~f.~Physik} {\bf C46} (1990) 63.

\bibitem{flynn:90}
{\sc J.~M. Flynn},
\newblock {\em Mod.~Phys.~Lett.} {\bf A5} (1990) 877.

\bibitem{albrechtetal:92}
{\sc {H.~Albrecht {\em et al.}~{\rm (HERA-B)}}},
\newblock {\em DESY preprint} {\bf DESY-PRC 92/04} (1992).

\bibitem{babar:93}
{\sc {\rm BaBar collaboration - Status Report}},
\newblock {\em SLAC preprint} {\bf SLAC-419} (June 1993).

\bibitem{nir:74}
{\sc Y.~Nir},
\newblock CP-Violation,
\newblock  {\bf SLAC-PUB-5874}.

\bibitem{burasetal:92d}
{\sc A.~J. Buras}, {\sc M.~Jamin}, and {\sc M.~E. Lautenbacher},
\newblock {\em Nucl.~Phys.} {\bf B408} (1993) 209.

\bibitem{ciuchini:92}
{\sc M.~Ciuchini}, {\sc E.~Franco}, {\sc G.~Martinelli}, and {\sc L.~Reina},
\newblock {\em Phys.~Lett.} {\bf 301B} (1993) 263.

\bibitem{buraslauten:93}
{\sc A.~J. Buras} and {\sc M.~E. Lautenbacher},
\newblock {\em Phys. Lett.} {\bf 318B} (1993) 212.

\bibitem{gibbons:93}
{\sc L.~K. Gibbons} {\em et~al.},
\newblock {\em Phys. Rev. Lett.} {\bf 70} (1993) 1203.

\bibitem{barr:93}
{\sc G.~D. Barr} {\em et~al.},
\newblock {\em Phys. Lett.} {\bf B317} (1993) 233.

\bibitem{burasetal:94a}
{\sc A.~J. Buras}, {\sc M.~E. Lautenbacher}, {\sc M.~Misiak}, and {\sc
  M.~M{\"u}nz},
\newblock {\em Nucl.~Phys.} {\bf B423} (1994) 349.

\bibitem{cohenetal:93}
{\sc A.~G. Cohen}, {\sc G.~Ecker}, and {\sc A.~Pich},
\newblock {\em Phys.~Lett.} {\bf 304B} (1993) 347.

\bibitem{pich:93}
{\sc A.~Pich},
\newblock  {\bf CERN-Th-7114-93} {\rm and refs.\ therein}.

\bibitem{harrisetal:93}
{\sc D.~A. Harris} {\em et~al.},
\newblock {\em Phys.~Rev.~Lett.} {\bf 71} (1993) 3918.
\end{thebibliography}
\end{document}